\documentclass[reprint,twocolumn,prb]{revtex4-2}
\usepackage{amsmath}
\usepackage{bm}
\usepackage{graphicx}
\usepackage{hyperref}
\usepackage{dcolumn}
\newcolumntype{d}[1]{D{.}{.}{#1}}

\begin{document}
\title{Strong shape-dependent intensity of inelastic light scattering by gold nanocrystals}

\author{Lucien Saviot}
\email{lucien.saviot@cnrs.fr}
\affiliation{
    Université Bourgogne Europe,
    CNRS,
    Laboratoire Interdisciplinaire Carnot de Bourgogne ICB UMR 6303,
    21000 Dijon,
    France
}

\author{Vincent Laude}
\affiliation{
    Universit\'e Marie et Louis Pasteur, CNRS, Institut FEMTO-ST,
    F-25000 Besançon,
    France
}

\begin{abstract}
We present a numerical approach to calculate inelastic light scattering spectra from gold nanocrystals, based on the finite element method.
This approach is validated by comparison with previous analytic calculations for spherically symmetric scatterers.
Superellipsoid nanocrystals are considered in order to smoothly vary the shape from octahedra to cubes via spheres, while preserving cubic symmetry.
Spectra are calculated and discussed taking into account the irreducible representation of the involved vibration modes.
A strong increase in the inelastically scattered light intensity is observed for small variations of the shape around the sphere.
This increase is related to variations of the electric field inside the nanocrystals, which are very small for small nanospheres but increase quickly for non-spherical nanocrystals.
This strong dependence with shape must be taken into account when interpreting experimental spectra acquired from inhomogeneous ensembles of nanocrystals whose shape dispersion are usually neglected.
The overall changes in the spectra when varying the shape of the nanocrystals provide additional insight into previously published results.
Preliminary calculations for chiral shapes further show a significant difference between spectra obtained with right or left circularly polarized light.
\end{abstract}

\maketitle

\section{Introduction}

The vibrations of objects are governed by their composition but also by their size, shape and environment.
Their exploration therefore provides a way to gain insight into all these properties.
At the nanoscale the frequencies of vibrations fall into the GHz to the THz range.
Different experimental methods have been used to detect them\cite{YuJPCC24}.
Inelastic light scattering by metallic nanoparticles has been shown to be suitable for this purpose more than 50 years ago\cite{GerstenPRB80}.
This experimental technique is still of interest nowadays thanks to continuous progress in the controlled synthesis of metallic nanocrystals and to the performance of Raman and Brillouin spectrometers.
Most of the existing literature in this domain relies on the assignment of modes to the peaks observed in the spectra, based on a comparison of experimental frequencies with calculated ones.
In this context, the Raman selection rules\cite{DuvalPRB92} for isotropic spheres have been very helpful in reducing the number of responsive vibrations.
However, this method becomes less and less useful as objects of lower symmetry are considered.
A few methods were proposed\cite{MontagnaPRB95,BachelierPRB04} in the past to calculate Raman spectra, but they have not been widely used, in particular because they have not been extended to non-spherically symmetric systems.
Recently, there has been a renewed interest in this domain with the appearance of an analytic approach for spherical nanoparticles composed of an elastically isotropic material\cite{GirardJCP17}, that was later extended to dimers\cite{GirardJPCC19}), and of other numerical approaches\cite{MontaOPriedePRB22,VasileiadisNL22,GelfandJPCC24}.
A few recent works have also been devoted to the related topic of numerical models of acousto-plasmonic coupling in metallic nanoparticles\cite{SaisonFranciosoJPCC20,OtomaloC22,BragasJOSAB23}.

The present work also aims at obtaining a general numerical approach to calculate the Raman spectrum for metallic nanoparticles of arbitrary shape but with additional constraints.
First, we want to demonstrate the validity of the numerical model by reproducing the spectra obtained analytically for spheres\cite{GirardJCP17}.
For this reason, we rely on the formalism developed in that work.
Second, we want to calculate spectra for nano-superellipsoids which have been shown to be a good approximation to the shape of rounded nanocubes, for which experimental spectra have been published in recent years\cite{TimmJPCC22,VernierAN23}.
By doing so, the shape can be smoothly varied from an octahedron to a cube via a sphere, preserving cubic symmetry and thus allowing the use of group theory\cite{SaviotN21} to help monitor the variations of the calculated spectra with shape.
Third, cubic elasticity is introduced in order to calculate spectra that are closer to experimental conditions.
Indeed, previous works have demonstrated that the splitting of the main low-frequency Raman peak is due to anisotropic elasticity in nanoparticles whose internal lattice structure is monodomain cubic gold or silver\cite{PortalesPNAS08,VernierAN23}.
In the following, we use the term ``nanocrystal'' in the latter case and ``nanoparticle'' otherwise.
Since the final goal is to devise a general method to calculate spectra for nanoparticles with arbitrary shapes, symmetry is not enforced during finite element calculations but used at the classification stage.
To interpret the results, free vibrations are indeed calculated using the Rayleigh-Ritz approach\cite{Visscher1991}.
This method takes advantage of the cubic symmetry\cite{Mochizuki1987,SaviotN21}.
It is very fast and allows determining the irreducible representations of the vibrations and to follow the dispersion of each mode as the shape is varied.

\section{Methods}
To calculate Raman spectra, we use the expressions derived by \citeauthor{GirardJCP17}\cite{GirardJCP17}.
They rely on a simple mechanism of Raman scattering, based on the density fluctuations in the
nanoparticles induced by the elastic vibrations.
Alternatives exist which also take into account the deformation potential coupling\cite{BachelierPRB04,SaisonFranciosoJPCC20,MontaOPriedePRB22,GelfandJPCC24}.
This is required when considering very small or non-metallic nanoparticles\cite{BachelierPRB04}.
This is not the case in this work and the deformation potential coupling can be neglected.

In this theory, the Stokes Raman intensity for each vibration eigenmode $i$ (pulsation $\omega_i$, displacement $\bm{u}_i$) of a homogeneous particle is given by overlap integrals involving the internal electric field $\bm{E}_\mathrm{int}$ and the vibration field, as
\begin{widetext}
\begin{equation}
  I_i(\omega)  \propto
    \frac{1+n(\omega)}{\omega}
    \left\|\bm{u} \times
      \left(\bm{u} \times
        \left(
          \iiint \bm{E}_\mathrm{int}(\bm{r}) e^{-i\bm{k}\bm{r}} \nabla\cdot\bm{u}_i(\bm{r}) dV
         -\iint \bm{E}_\mathrm{int}(\bm{r}) \left(\bm{u}_i(\bm{r})\cdot\bm{n}\right) dS
        \right)
      \right)
    \right\|^2 \delta(\omega-\omega_i)
    \label{eqI}
\end{equation}
\end{widetext}
The volume and surface integrals in Eq.~(\ref{eqI}) are over the nanoparticle only, $n(\omega)$ is the Bose factor, $\bm{k}=\|\bm{k}\| \bm{u}$ is the scattered wave-vector, and $\bm{n}$ is the outward normal to the surface.
As a first-order perturbation theory, acousto-plasmonic coupling is evaluated from the unperturbed internal electric field and nanoparticle vibration, in contrast to a full-wave approach that requires solving the scattered electric field for every particle vibration\cite{SaisonFranciosoJPCC20}.
Equation ~(\ref{eqI}) is further reminiscent of optomechanical coupling in dielectric nanostructures\cite{PennecN14}, that also involves a linear combination of volume and surface overlap integrals.

$\bm{E}_\mathrm{int}$ and $\bm{u}_i$ are both calculated using the finite element method (FEM) with FreeFem++\cite{freefem}, as in Ref.~\onlinecite{SaisonFranciosoJPCC20}.
Vibration eigenmodes are obtained by solving the Navier equation for a nanoparticle defined by its mass density $\rho$, elasticity tensor $C_{IJ}$, and shape.
We neglect the embedding matrix in this calculation so that only free vibrations are considered.
This assumption is generally valid, in particular when the matrix mass density and the coefficients of the elasticity tensor are smaller than those of the nanoparticle\cite{MurrayPRB04}.
Should radiation of elastic waves inside the host medium be included, it would be required to properly account for leakage, for instance using the concept of quasi-normal modes\cite{LaudePRB2023}, but this goes beyond the scope of the present work.

For the electric field calculation, we consider the plane-wave illumination of the single metallic nanoparticle at rest, with permittivity $\epsilon$, embedded in an homogeneous dielectric medium defined by its real permittivity $\epsilon_m = n_m^2$.
We solve the vector wave equation as detailed in Ref.~\onlinecite{SaisonFranciosoJPCC20}.

\begin{figure}
    \includegraphics[width=\columnwidth]{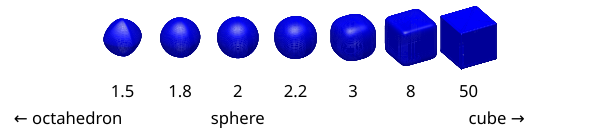}
    \caption{\label{sq}Superellipsoids of cubic symmetry considered in this work. The shape varies from an octahedron for $n=1$ (left side) to a cube as $n \to \infty$ (right side), via a sphere at $n=2$.}
\end{figure}

We consider superellipsoid (or superquadrics) nanoparticles whose surface is defined by the implicit equation
\begin{equation}
    \left|\frac{x}{L}\right|^n +
    \left|\frac{y}{L}\right|^n +
    \left|\frac{z}{L}\right|^n = 1.
    \label{sqeq}
\end{equation}
Their shape varies from octahedra for $n=1$ to cubes for $n \to \infty$ via spheres for $n=2$.
Their length in the $x$, $y$ and $z$ directions is $2L$.
The shapes considered in this work are represented in Fig.~\ref{sq}.

Superellipsoids are meshed with gmsh\cite{gmsh}, starting from the parametric description of their surface
\begin{eqnarray}
    \label{parametric}
    x(t,p) & = & L \; c(t,n) \; c(p,n),\nonumber\\
    y(t,p) & = & L \; c(t,n) \; s(p,n),\\
    z(t,p) & = & L \; s(t,n),\nonumber
\end{eqnarray}
with $0 \le t \le \pi$ and $0 \le p \le 2\pi$.
Coordinates $t$ and $p$ identify with the usual spherical coordinates $\theta$ and $\phi$ for $n=2$ only.
Functions $c(x,n)$ and $s(x,n)$ are defined as
\begin{eqnarray}
    c(x,n) & = & \cos{x} \; |\cos{x}|^{2/n-1} ,\nonumber\\
    s(x,n) & = & \sin{x} \; |\sin{x}|^{2/n-1} .
\end{eqnarray}

\section{Results and Discussion}

\subsection{Spherical nanoparticles}

\begin{figure}
    \includegraphics[width=0.8\columnwidth]{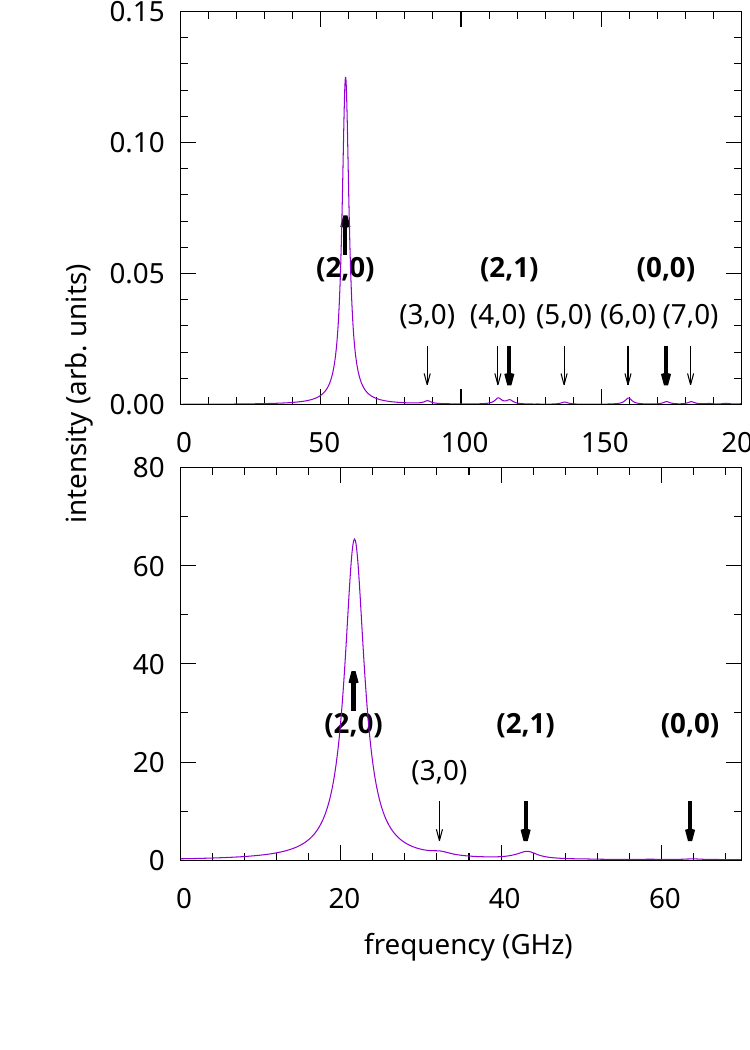}
    \caption{\label{JLFig3}Raman spectra computed for two isotropic gold spheres embedded in a transparent matrix of index $n_m=1.5$, for a 520.8~nm excitation wavelength and in the backscattering geometry.
    Two diameters are considered, 18~nm (top) and 49~nm (bottom).
    Labels $(\ell,n)$ and arrows indicate the frequencies of the $n$\textsuperscript{th} overtone of the spheroidal mode with angular momentum $\ell$.
    Raman active modes are indicated in bold.
    Compare with Fig.~3 of Ref.~\onlinecite{GirardJCP17}}.
\end{figure}

We first apply the method presented above to the case of isotropic spheres, in order to reproduce the analytic calculations from \citeauthor{GirardJCP17}\cite{GirardJCP17}.
Fig. \ref{JLFig3} shows spectra for isotropic gold spheres with diameters 18 and 49~nm that can be directly compared with those presented in Fig.~3 of Ref.~\onlinecite{GirardJCP17}.
The elastic parameters for gold are obtained from the sound velocities ($v_L=3330$ and $v_T=1250$~m/s). The mass density is $\rho=19300$~kg/m$^3$.
Optical constants are taken from \citeauthor{JohnsonPRB72}\cite{JohnsonPRB72}.
For simplicity, we used $\lambda=520.8$~nm ($\epsilon = (0.62 + 2.081 i)^2$).
That wavelength is close to $\lambda=532$~nm and no significant change is expected from this minor difference.
The nanospheres are embedded in a transparent matrix of index $n_m=1.5$.
Spectra are calculated for the back-scattering configuration.
The 300 lowest frequency vibrations are taken into account.
They cover the frequency range up to about 1.4 times the breathing mode frequency (fundamental spheroidal mode with $\ell=0$).
The different contributions were summed after broadening with a Lorentzian shape function with a full width at half-maximum of 3 GHz, as in Ref.~\onlinecite{GirardJCP17}.

The agreement with the original figure is very good for the larger nanoparticle, for which most vibrations have a negligible contribution to the spectrum, except for the expected Raman active ones: fundamental and first overtone of the quadrupolar mode (spheroidal $\ell=2$), and a minor contribution from the fundamental breathing mode (spheroidal $\ell=0$).
The weak spheroidal $\ell=3$ peak also agrees with the analytic calculations.
It results from the relaxation of the selection rules for nanoparticles that are not very small compared to the optical wavelength.

The agreement is good for the smaller nanoparticle as well, except for the additional weak contributions from spheroidal modes with $\ell>2$.
As already pointed out in Ref.~\onlinecite{GirardJCP17}, the intensity of the main peak is in this case three orders of magnitude lower than for the larger nanoparticle.
This fact results from the almost exact cancellation of volume and surface integrals in Eq.~(\ref{eqI}).
The spurious peaks have even smaller intensities.
Reproducing such small intensities is challenging with numerical approaches.
The spherical shape is imperfectly modeled due to the mesh discretization and to the convergence of the finite element method.
The results reported in Fig.~\ref{JLFig3} are therefore promising, since all the main features of the calculated spectra are reproduced, except for minor deviations in the most numerically challenging cases.

\subsection{Superellipsoidal nanoparticles}

\begin{figure*}
    \includegraphics[width=0.75\textwidth]{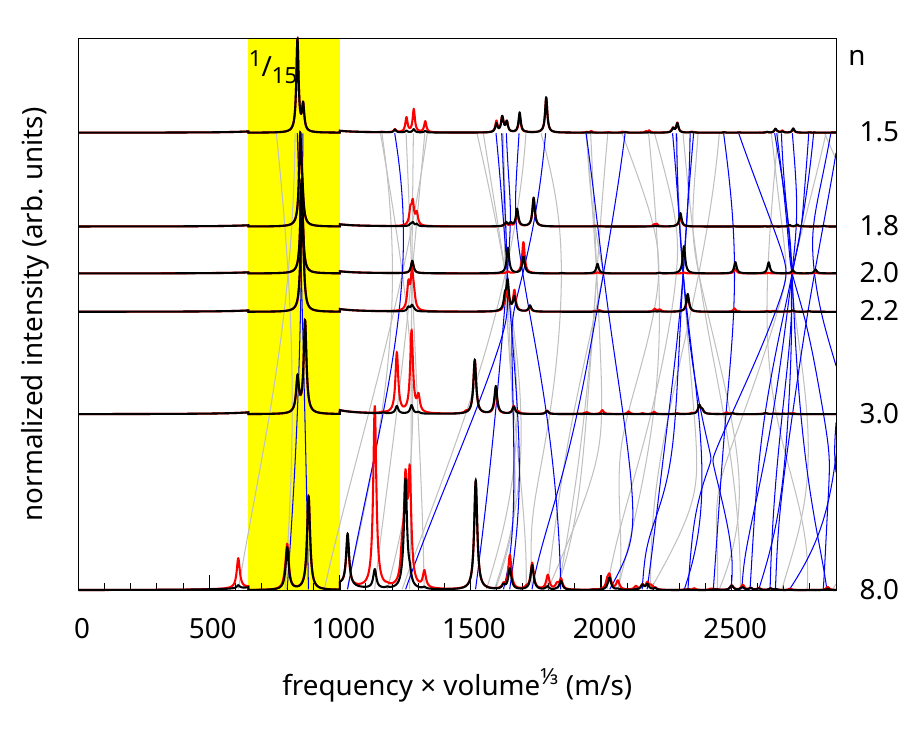}
    \caption{\label{shape}Raman spectra computed for isotropic gold superellipsoid nanoparticles with $2L=18$ (black) and 49~nm (red, below) and $n=1.5$, 1.8, 2, 2.2, 3 and 8 from top to bottom.
    Spectra are normalized by the maximum intensity and shifted vertically as $1/n$ for
    clarity and to overlay frequency variations with $n$.
    Intensities are divided by a factor $15$ for reduced frequencies ranging from 650 to 1100~m/s.
    The blue and gray curves show the dispersion curves for Raman active (A\textsubscript{1g}, E\textsubscript{g} and T\textsubscript{2g}) and inactive vibrations, respectively.}
\end{figure*}

To demonstrate the usefulness of the present approach, we now consider the case of non-spherically symmetric particles that can not be modeled analytically.
To do so, we first vary the shape of the nanoparticles.
Fig.~\ref{shape} shows normalized spectra calculated for different superellipsoids chosen so that the shape varies around the sphere while keeping $2L=18$ or 49~nm.
Frequencies are multiplied by the cubic root of the volume and expressed as sound velocities, as in previous works\cite{TimmJPCC22,VernierAN23}.
Using these reduced frequencies, the position of Raman active vibrations remain mostly unchanged while varying $L$ and $n$\cite{SaviotN21}.
The peaks are now broadened with narrower Lorentzian functions to better highlight individual contributions.
Vibrations are also calculated using the Rayleigh-Ritz method\cite{SaviotN21} for each irreducible representations of O\textsubscript{h} with varying $n$.
The resulting dispersion curves are added to Fig.~\ref{shape} for both Raman active (blue) and inactive (grey) modes to help assign peaks.
The spectra contain two intense E\textsubscript{g} and T\textsubscript{2g} peaks that merge into the quadrupolar vibrations of the sphere ($n=2$) at 853~m/s.
This modal splitting results from cubic symmetry ($n \neq 2$) being lower than spherical symmetry.
The incident light polarization is kept along a (100) direction and the E\textsubscript{g} peak is more intense.
Additional weaker peaks at higher frequency are present even for $2L=18$~nm when $n \ne 2$.
Their assignment will be discussed later.

\begin{figure}
    \includegraphics[width=0.9\columnwidth]{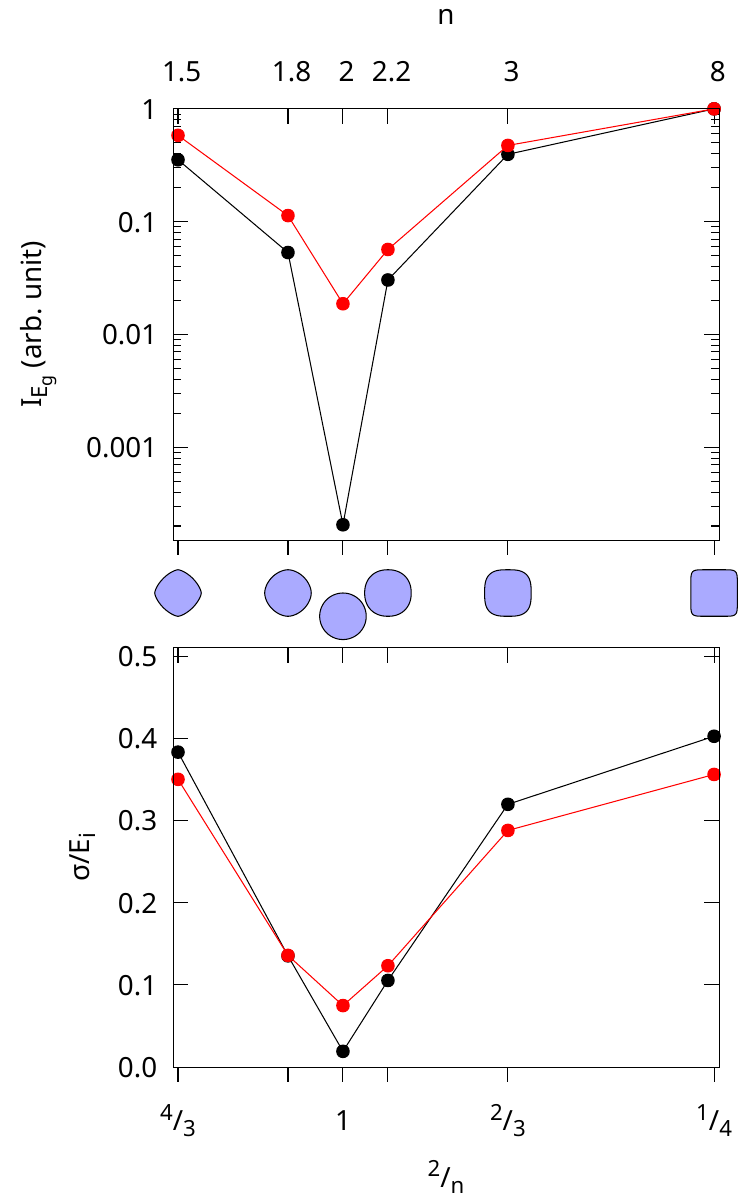}
    \caption{\label{Ishape}Intensity of the E\textsubscript{g} band normalized at $n=8$ (top) and standard deviation $\sigma$ of the electric field amplitude inside the nanoparticle normalized to the amplitude of the incident electric field (bottom), for different superellipsoids whose cross-section along the $x=0$ plane is represented in the middle.}
\end{figure}

\begin{figure}
    \includegraphics[width=\columnwidth]{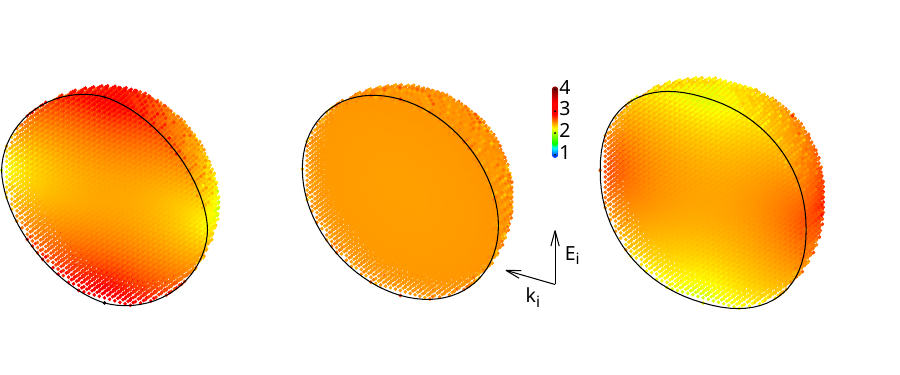}
    \caption{\label{Ein}$\|\bm{E_{int}}\|$ inside gold superellipsoids with $2L=18$~nm for $n=1.8$, 2 and 2.2 from left to right. The directions of the incident field $\bm{E_i}$ and propagation $\bm{k_i}$ are indicated with arrows.}
\end{figure}

The most striking point regarding these spectra is their intensity that increases sharply when the shape deviates from the sphere, as seen in Fig.~\ref{Ishape} (top).
This variation is strong, in particular for the smaller nanoparticles for which the intensity increases by two orders of magnitude between $n=2$ and $n=2.2$ or $n=1.8$ while the volume variation remains small ($\pm10\%$).
\citeauthor{GirardJCP17} pointed out that the main ingredient driving the Raman intensity is the spatial variation of the electric field inside the nanoparticle.
Therefore, to understand the origin of the large intensity variations, we plot the magnitude of the electric field inside the $n=1.8$, 2 and 2.2 nanoparticles with $2L=18$~nm in Fig.~\ref{Ein}.
The internal electric field is almost the same everywhere inside the sphere.
This is expected because it is exactly constant for a dielectric sphere in the electrostatic case (an analytic derivation is provided in Ref.~\onlinecite{Jackson}).
It is not exactly zero in the numerical calculations because the retardation effect is taken into account (variation of the phase of the incident electric field along the propagation direction) as in Mie scattering theory.
The average value is nevertheless close to $\left|\frac{3\epsilon_m}{\epsilon+\epsilon_m} E_{inc}\right|$ for $2L=18$~nm.
We calculated the standard deviation $\sigma$ of the magnitude of the electric field inside the nanoparticle to quantify its variation.
The results plotted in Fig.~\ref{Ishape} (bottom) show that $\sigma$ varies quickly around $n=2$.
When $n$ varies from 2 to 1.8 for $2L=18$~nm, the intensity increases by two orders of magnitude while the volume decreases.
To reach the same intensity, it would be necessary to multiply the volume $V$ of the $2L=18$~nm sphere by $\simeq 9$ according to the $V^{2.3}$ intensity variation obtained in Ref.~\onlinecite{GirardJCP17}.
The difference is even more striking when the shape tends to an octahedron (larger intensity increase as the volume becomes smaller).
Similarly, when going from a sphere to a cube while keeping $L$ constant, the volume is approximately doubled.
The $V^{2.3}$ law predicts that the intensity is multiplied by $\simeq 5.5$ which is much less than the 2 or 4 orders of magnitude observed in Fig.~\ref{Ishape} (top).
From this point of view, the variation of intensity with shape is significantly stronger than the variation with size.

To the best of our knowledge, this strong shape dependence has not been reported previously in the literature.
Yet the spherical shape approximation is never perfectly valid, because actual nano-objects are made of a finite number of atoms.
In addition, nanocrystals are often faceted as can be predicted for example by Wulff construction.
It is therefore necessary to be careful when comparing experimental results with the spherical approximation, in particular when discussing the intensities for small, almost spherical nanoparticles.

\subsection{Symmetry and Raman active modes}

We now turn to the modal assignment of the lower intensity features in the spectra.
As explained before, care should be taken for small spheres due to the very low intensity of the spectrum resulting in the appearance of spurious peaks due to numerical and discretization issues.
Thanks to the strong intensity increase discussed before, this point is less problematic for non-spherical nanoparticles.
Peaks around 1700~m/s in Fig.~\ref{shape} are present in most spectra.
The corresponding vibrations are Raman active modes (A\textsubscript{1g}, E\textsubscript{g} and T\textsubscript{2g}) coming from the (2,1) and (4,0) spheroidal modes of the sphere.
For the larger nanoparticles, peaks are also present around 1300~m/s and come from the previously mentioned (3,0) spheroidal mode of the sphere.
For the shape closest to a cube ($n=8$), additional peaks are observed near 610 and 1130, 1250 and 1320~m/s.
The first one comes from the (2,0) torsional modes, whereas the others come from the (3,0) spheroidal and (3,0) torsional modes of the sphere.
Raman scattering by torsional modes is generally considered to be negligible, including for large nanoparticles.
This is because such vibrations do not modify the mass density during oscillation, but also because the shape of a sphere does not change during oscillations caused by torsional vibrations.
It is indeed well-known for spheres that the Raman intensity is associated to the ability of the vibration to change the shape during oscillation\cite{BachelierPRB04}.
For cubes, torsional vibrations modify the shape, which explains why the Raman scattering cross-sections for these vibrations becomes non-negligible.
However, contrarily to the strong shape dependence reported above for the intensity, the apparition of peaks due to torsional modes is only observed for values of $n$ significantly different from 2.

\subsection{Superellipsoidal nanocrystals}

A detailed comparison with experiments is out of the scope of the present work because
results reported in the literature differ in nanocrystal composition, shape, size, excitation wavelength, optical index of the surrounding medium, different ligands, presence of neighbouring nanocrystals, \ldots{}
In the following, we highlight some general features of the Raman spectra of metallic nanocubes and their manifestation in published spectra.
While the previous calculations provide an interesting insight in Raman scattering from non-spherically symmetric nanoparticles, it does not take elastic anisotropy into account.
Gold and silver nanocubes are often single domain (not twinned) and their shape comes from the cubic lattice structure of gold.
This has to be taken into account to model the vibration frequencies of gold nanocrystals accurately and assign the observed Raman peaks\cite{PortalesPNAS08}.
For this reason, we calculated the spectra for the same $2L=49$~nm gold nanocrystals using the elastic parameters for cubic gold ($\rho=19.283$~g/cm$^3$, $C_{11}=191$, $C_{12}=162$ and $C_{44}=42.4$~GPa).
The $x$, $y$ and $z$ axes of the superellipsoids are along the ${<}100{>}$ directions of the lattice structure.

\begin{figure*}
    \includegraphics[width=0.75\textwidth]{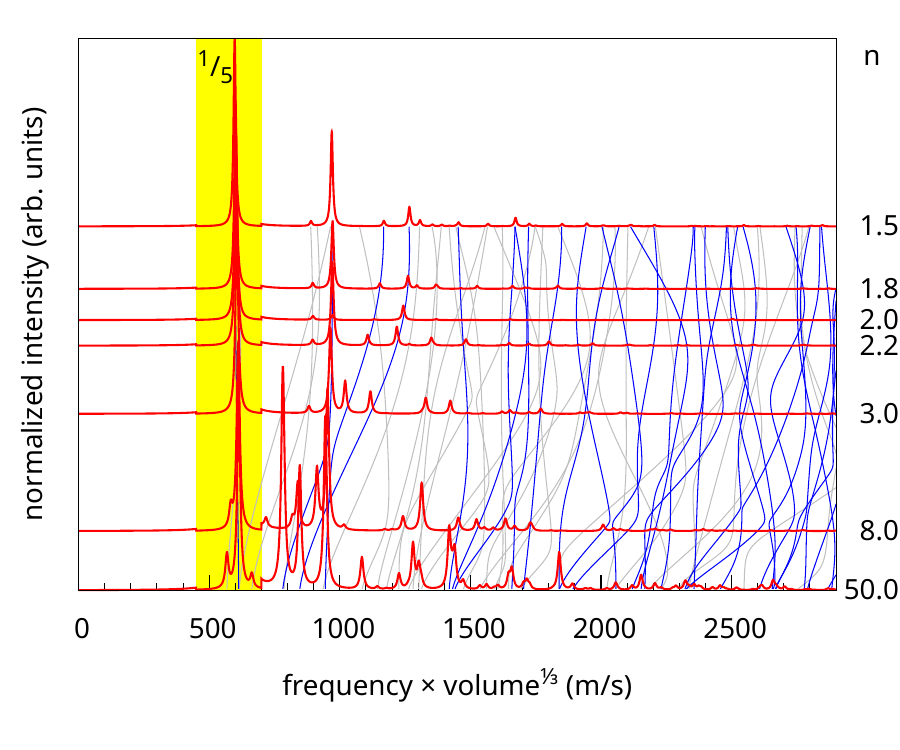}
    \caption{\label{shape-ani}Raman spectra computed for cubic Au superellipsoid nanocrystals $2L=49$~nm (red) and $n=1.5$, 1.8, 2, 2.2, 3, 8 and 50 from top to bottom.
    The spectra are normalized at the maximum intensity and shifted vertically as $\frac1n$ for clarity
    and to overlay the frequency variations with $n$.
    Intensities are divided by 15 between 450 and 700~m/s.
    The blue and grey curves show the normalized frequencies of the Raman active (A\textsubscript{1g}, E\textsubscript{g} and T\textsubscript{2g}) and inactive vibrations respectively.}
\end{figure*}

Computed normalized spectra are plotted in Fig.~\ref{shape-ani}, with the same conventions as in Fig.~\ref{shape}.
Note that there is a strong relationship between the intensities calculated for nanocrystals with identical shapes but different elasticity tensors.
Indeed, the displacements associated to each vibration mode for each system form an orthogonal basis.
The vibrations of one system are therefore fully described by projecting onto the second system\cite{SaviotPRB09}.
Therefore, the strong shape dependence of the intensity discussed before for isotropic elasticity is still valid for cubic gold.
As a note, the Raman spectrum for a sphere made of an elastically anisotropic material can be obtained more accurately by projecting the vibrations of the anisotropic sphere obtained numerically (FEM or Rayleigh-Ritz for example) onto those of the isotropic sphere (Lamb modes\cite{LambPLMS1881}) and calculating the intensities using the analytic approach\cite{GirardJCP17}.

The main Raman peaks in Fig.~\ref{shape-ani} correspond to the E\textsubscript{g} and T\textsubscript{2g} peaks at about 600 and 950~m/s, coming from the splitting of the intense spheroidal $\ell=2$ peak of the isotropic sphere.
The spectra also show minor contributions to the high frequency side of the T\textsubscript{2g} peak which shift to the low frequency side for $n>3$.
A similar behavior was recently reported for silver nanocubes\cite{VernierAN23}.
It was assigned to the anti-crossing between two T\textsubscript{2g} branches.
The present calculations show an additional contribution of the second E\textsubscript{g} branch.
Interesting experimental Raman spectra for gold nanocrystals were also reported recently\cite{TimmJPCC22}.
They showed very narrow peaks because they were obtained for single nanocrystals and therefore without inhomogeneous broadening, contrarily to most reported results involving ensemble measurements,\textit{i.e.}, for which the spectra contain the contribution of several nanocrystals.
The present calculations, and in particular those presented in Fig.~\ref{shape-ani}, shed a new light on these results.
The substructures observed for the E\textsubscript{g} and T\textsubscript{2g} peaks were tentatively assigned to their splitting due to unequal dimensions of the nanocrystals along the 3 directions.
Even so, the spectrum near the expected E\textsubscript{g} peak seemed to be composed of 3 contributions which could not be explained by the two-fold degeneracy of the E\textsubscript{g} mode.
The spectra calculated for $n=50$ show that additional Raman inactive vibrations can contribute in this frequency range.
Their calculated intensity is here small, but the nanocubes in Ref.~\onlinecite{TimmJPCC22} were larger ($2L\sim80$~nm).
The large shift observed for the peak near the T\textsubscript{2g} frequency when changing the incident polarization was also not explained.
The present calculations show that additional contributions in this frequency range are indeed possible, as discussed before for silver nanocubes.
Finally, the nanocrystals showing the most unexpected features in Ref.~\onlinecite{TimmJPCC22} were those close to a silica step.
In addition, the spectra were sensitive to the polarization of the incident light.
The presence of this step was also shown to modify the electric field magnitude inside the nanocube.
Different incident polarizations are expected to modify $\bm{E}_\mathrm{int}$.
As detailed above, these modifications can in principle change the calculated spectra.
Taking into account all these factors goes beyond the scope of this work, but will be considered in the future.
They may open the door to monitoring the electric field inside metallic nanoparticles through inelastic scattering measurements.

As indicated in the introduction, \citeauthor{MontaOPriedePRB22}\cite{MontaOPriedePRB22}  recently reported similar calculations for small gold spherical nanocrystals made of isotropic and cubic gold.
As discussed before, calculations for anisotropic spheres can be handled analytically after projection of the vibrations onto those of an isotropic sphere.
We note also that the method used in that work requires the calculation of the electric field for the sphere deformed by each vibration mode, which requires significantly more computations.
Calculations were actually performed only for a few selected vibrations (spheroidal $\ell=0$ and 2).
This selection is valid for an isotropic sphere, but questionable for an anisotropic sphere whose vibrations are not pure Lamb modes anymore; it can not be applied non-spherical shapes.

In another recent work, \citeauthor{GelfandJPCC24}\cite{GelfandJPCC24} presented calculated spectra for small isotropic silver and gold spheres and a silver cube.
Similar comments regarding the analytical solution for spheres and the larger computational resources required to calculate spectra are valid.
The authors did not attempt to check the validity of the Raman selection rules, arguing that Raman inactive vibrations have been experimentally observed in Ref.~\onlinecite{KuokPRL03}.
However, the cited work concerns much larger silica spheres with diameters $\ge200$~nm.
The Raman selection rules are indeed invalid in that case because they are derived for sizes that are small compared to the wavelength of light.
The authors also argue that the Raman selection rules break down for non-totally symmetric vibrations because the deformed nanoparticle does not have the same symmetry as the undeformed one.
This reasoning seems dubious because it should also apply to Raman scattering in molecules.
It is difficult to discuss in detail here the results presented by the authors because the peaks in the spectra of spheres were not assigned to Lamb modes.
However, we note that the relative intensities for the silver sphere are quite different from those reported in Ref.~\onlinecite{BachelierPRB04}.
While the reason for this discrepancy is unknown, it may be related to a poor convergence of the calculations in this challenging case, as explained before.
We also believe that our work demonstrates that the Raman selection rules remain useful to follow the evolution of the intensity of each mode, in particular when smoothly varying the shape of the nanoparticles.
A direct comparison with spectra calculated in \citeauthor{VasileiadisNL22}\cite{VasileiadisNL22} is difficult since that work focused on nanorods which are not considered in the present work.
We note however that the electromagnetic coupling in dimers was shown to play a significant role, which agrees with the sensitivity to the variation of the electric field discussed in the present work.

\subsection{Chiral nanoparticles}

\begin{figure}
    \includegraphics[width=\columnwidth]{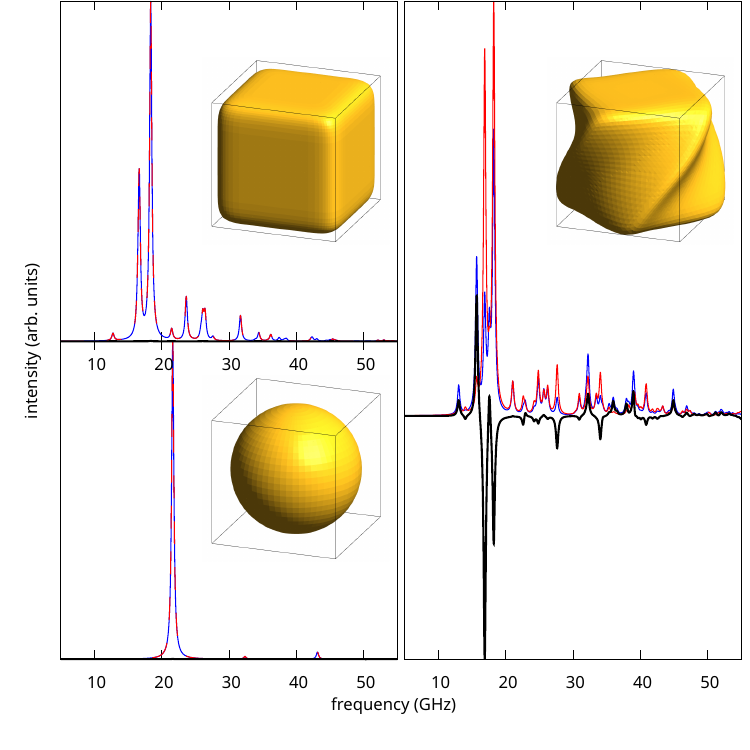}
    \caption{\label{ROA}Raman spectra for left (blue line) and right (red line) circularly polarized incident light, and their difference (black line) for a sphere (bottom left panel), a rounded cube with $n=8$ (top left panel) and a rounded twisted cube (right panel). The 3D views of the nanoparticles are plotted within a cube with edge length $2L$. The incident beam propagates in the direction perpendicular to the front face of the cubes. The rounded twisted cube is obtained from the rounded cube by applying a rotation along the vertical axis $z$ by an angle varying linearly with $z$ so that the top and bottom faces are rotated by $\pm45^\circ$.}
\end{figure}

We now turn to calculations for chiral gold nanoparticles excited with circularly polarized light.
To the best of our knowledge, no such measurements have been reported in the literature despite the strong current interest in chirality.
These calculations are motivated by the similarity with the probing of molecular chirality using Raman optical activity (ROA)\cite{BarronJMS07}.
We thus aim at investigating the possibility of probing chirality in gold nanoparticles.
To this end, we consider chiral nanoparticles that are identical or as close as possible to the previous ones, to ease comparison.
We calculated spectra for right (clockwise from the point of view of the receiver) and left circularly polarized incident light for a sphere, a rounded cube and a twisted rounded cube made of gold with isotropic elasticity.
We used the same length ($2L=49$~nm) for all of them.
Only the last particle has a chiral shape.
It is obtained starting from the rounded cube (superellipsoid with $n=8$) by rotating the points around the $z$-axis by the $z$-dependent angle $\alpha(z)=\frac{\pi}{4}\frac{z}{L}$.

The calculated spectra and 3D representations of the nanoparticles are presented in Fig.~\ref{ROA}.
The volume of the twisted and non-twisted rounded cube are the same, so their vibrational frequencies are close.
Indeed, the peaks with the largest intensity fall in the same frequency range.
As expected, the spectra calculated for the right and left circular polarizations are equal (within numerical precision) for achiral nanoparticles (sphere and rounded cube).
A large difference is obtained, in contrast, for the twisted rounded cube.
ROA is a very small effect with a magnitude of about $10^{-3}$ or less requiring special detection schemes to observe it.
In the present case, the magnitude is larger than $10^{-1}$, suggesting that observation may be possible using existing setups.
Indeed, even if all the peaks near the quadrupolar-like vibration would merge when adding an inhomogeneous broadening, as in ensemble measurements for example, the position of this peak would shift between the two circular polarization because the sign of the difference is opposite for the low and high frequency sides.
We assign this larger magnitude to the large size of the nanoparticles compared to molecules, which makes it easier to observe an effect due to the spatial variation of the incident electric field, but also to the fact that we consider the vibrations of a larger entity (compared to the vibrations a few molecular bonds in the Raman spectra of molecules).
The validity and the applicability of these calculations must of course be reevaluated when investigating real systems.
In particular, the shape we considered is somewhat arbitrary, even if more complex shapes have been reported in the literature\cite{ChoAN20}.
The environment of the nanoparticle can also play a significant role and inhomogeneous broadening in ensemble measurements may further hinder this difference.

\section{Conclusion}
A numeric approach based on the finite element method has been proposed to calculate inelastic light scattering spectra of metallic nanoparticles.
The approach satisfactorily reproduces the results of the analytical model from which it was derived.
Calculations for superellipsoids have revealed a strong variation of the intensity of scattered light with shape, in particular for small nanoparticles.
This effect originates from the special case of spheres for which the internal electric field is constant in the electrostatic approximation, resulting in a very small intensity originating only from small retardation effects.
Going away from this particular case by changing the shape is shown to increase the inelastically scattered intensity by orders of magnitude.
This point must carefully be taken into account when the scattering intensity plays a role, as in ensemble measurements for example.
Indeed, within an inhomogeneous population of almost spherical nanoparticles, the less spherical ones may contribute more to the spectrum, making calculations based on the spherical approximation misleading.
Spectra calculated for nanocubes with cubic elasticity show the contribution of vibration modes which have not been considered before to assign experimental features.
In particular, this work supports the interpretation of original features observed experimentally on single gold nanocubes as originating from the variation of the internal electric field inside the nanocubes when modifying the incident polarization or the environment, as observed for nanocrystals near a silica step.
This opens the door to using inelastic light scattering as an indirect probe of the inner electric field in metallic nano-objects.
Finally, preliminary calculations show that the spectra of chiral nano-objects with right and left circularly polarized incident light differ significantly making inelastic light scattering spectroscopy suitable to assess the chirality of nano-objects.

\begin{acknowledgments}
This work has been supported by the EIPHI Graduate School (contract ANR-17-EURE-0002) operated by the French National Research Agency (ANR).
L.S. acknowledges support by ANR CHIRNATIO (ANR-22-CE09-0007) and thanks Oph\'elie Saison-Francioso for providing a working FreeFEM script to calculate the electric field using second-order N\'ed\'elec tetrahedral finite element.
Calculations were performed using HPC resources from DNUM CCUB (Centre de Calcul de l'Universite\'e de Bourgogne).

\end{acknowledgments}

\bibliography{Isq}

\begin{thebibliography}{31}%
\makeatletter
\providecommand \@ifxundefined [1]{%
 \@ifx{#1\undefined}
}%
\providecommand \@ifnum [1]{%
 \ifnum #1\expandafter \@firstoftwo
 \else \expandafter \@secondoftwo
 \fi
}%
\providecommand \@ifx [1]{%
 \ifx #1\expandafter \@firstoftwo
 \else \expandafter \@secondoftwo
 \fi
}%
\providecommand \natexlab [1]{#1}%
\providecommand \enquote  [1]{``#1''}%
\providecommand \bibnamefont  [1]{#1}%
\providecommand \bibfnamefont [1]{#1}%
\providecommand \citenamefont [1]{#1}%
\providecommand \href@noop [0]{\@secondoftwo}%
\providecommand \href [0]{\begingroup \@sanitize@url \@href}%
\providecommand \@href[1]{\@@startlink{#1}\@@href}%
\providecommand \@@href[1]{\endgroup#1\@@endlink}%
\providecommand \@sanitize@url [0]{\catcode `\\12\catcode `\$12\catcode
  `\&12\catcode `\#12\catcode `\^12\catcode `\_12\catcode `\%12\relax}%
\providecommand \@@startlink[1]{}%
\providecommand \@@endlink[0]{}%
\providecommand \url  [0]{\begingroup\@sanitize@url \@url }%
\providecommand \@url [1]{\endgroup\@href {#1}{\urlprefix }}%
\providecommand \urlprefix  [0]{URL }%
\providecommand \Eprint [0]{\href }%
\providecommand \doibase [0]{https://doi.org/}%
\providecommand \selectlanguage [0]{\@gobble}%
\providecommand \bibinfo  [0]{\@secondoftwo}%
\providecommand \bibfield  [0]{\@secondoftwo}%
\providecommand \translation [1]{[#1]}%
\providecommand \BibitemOpen [0]{}%
\providecommand \bibitemStop [0]{}%
\providecommand \bibitemNoStop [0]{.\EOS\space}%
\providecommand \EOS [0]{\spacefactor3000\relax}%
\providecommand \BibitemShut  [1]{\csname bibitem#1\endcsname}%
\let\auto@bib@innerbib\@empty
\bibitem [{\citenamefont {Yu}\ and\ \citenamefont {Wang}(2024)}]{YuJPCC24}%
  \BibitemOpen
  \bibfield  {author} {\bibinfo {author} {\bibfnamefont {K.}~\bibnamefont
  {Yu}}\ and\ \bibinfo {author} {\bibfnamefont {G.~P.}\ \bibnamefont {Wang}},\
  }\bibfield  {title} {\bibinfo {title} {Coherent acoustic vibrational spectrum
  of single and coupled nanoresonators},\ }\href
  {https://doi.org/10.1021/acs.jpcc.4c00256} {\bibfield  {journal} {\bibinfo
  {journal} {J. Phys. Chem. C}\ }\textbf {\bibinfo {volume} {128}},\ \bibinfo
  {pages} {5394} (\bibinfo {year} {2024})}\BibitemShut {NoStop}%
\bibitem [{\citenamefont {Gersten}\ \emph {et~al.}(1980)\citenamefont
  {Gersten}, \citenamefont {Weitz}, \citenamefont {Gramila},\ and\
  \citenamefont {Genack}}]{GerstenPRB80}%
  \BibitemOpen
  \bibfield  {author} {\bibinfo {author} {\bibfnamefont {J.~I.}\ \bibnamefont
  {Gersten}}, \bibinfo {author} {\bibfnamefont {D.~A.}\ \bibnamefont {Weitz}},
  \bibinfo {author} {\bibfnamefont {T.~J.}\ \bibnamefont {Gramila}},\ and\
  \bibinfo {author} {\bibfnamefont {A.~Z.}\ \bibnamefont {Genack}},\ }\bibfield
   {title} {\bibinfo {title} {Inelastic {M}ie scattering from rough metal
  surfaces: Theory and experiment},\ }\href
  {https://doi.org/10.1103/physrevb.22.4562} {\bibfield  {journal} {\bibinfo
  {journal} {Phys. Rev. B}\ }\textbf {\bibinfo {volume} {22}},\ \bibinfo
  {pages} {4562} (\bibinfo {year} {1980})}\BibitemShut {NoStop}%
\bibitem [{\citenamefont {Duval}(1992)}]{DuvalPRB92}%
  \BibitemOpen
  \bibfield  {author} {\bibinfo {author} {\bibfnamefont {E.}~\bibnamefont
  {Duval}},\ }\bibfield  {title} {\bibinfo {title} {Far-infrared and {R}aman
  vibrational transitions of a solid sphere: Selection rules},\ }\href
  {https://doi.org/10.1103/physrevb.46.5795} {\bibfield  {journal} {\bibinfo
  {journal} {Phys. Rev. B}\ }\textbf {\bibinfo {volume} {46}},\ \bibinfo
  {pages} {5795} (\bibinfo {year} {1992})}\BibitemShut {NoStop}%
\bibitem [{\citenamefont {Montagna}\ and\ \citenamefont
  {Dusi}(1995)}]{MontagnaPRB95}%
  \BibitemOpen
  \bibfield  {author} {\bibinfo {author} {\bibfnamefont {M.}~\bibnamefont
  {Montagna}}\ and\ \bibinfo {author} {\bibfnamefont {R.}~\bibnamefont
  {Dusi}},\ }\bibfield  {title} {\bibinfo {title} {Raman scattering from small
  spherical particles},\ }\href {https://doi.org/10.1103/PhysRevB.52.10080}
  {\bibfield  {journal} {\bibinfo  {journal} {Phys. Rev. B}\ }\textbf {\bibinfo
  {volume} {52}},\ \bibinfo {pages} {10080} (\bibinfo {year}
  {1995})}\BibitemShut {NoStop}%
\bibitem [{\citenamefont {Bachelier}\ and\ \citenamefont
  {Mlayah}(2004)}]{BachelierPRB04}%
  \BibitemOpen
  \bibfield  {author} {\bibinfo {author} {\bibfnamefont {G.}~\bibnamefont
  {Bachelier}}\ and\ \bibinfo {author} {\bibfnamefont {A.}~\bibnamefont
  {Mlayah}},\ }\bibfield  {title} {\bibinfo {title} {Surface plasmon mediated
  {R}aman scattering in metal nanoparticles},\ }\href
  {https://doi.org/10.1103/physrevb.69.205408} {\bibfield  {journal} {\bibinfo
  {journal} {Phys. Rev. B}\ }\textbf {\bibinfo {volume} {69}},\ \bibinfo
  {pages} {205408} (\bibinfo {year} {2004})}\BibitemShut {NoStop}%
\bibitem [{\citenamefont {Girard}\ \emph {et~al.}(2017)\citenamefont {Girard},
  \citenamefont {Lerm\'e}, \citenamefont {Gehan}, \citenamefont {Margueritat},\
  and\ \citenamefont {Mermet}}]{GirardJCP17}%
  \BibitemOpen
  \bibfield  {author} {\bibinfo {author} {\bibfnamefont {A.}~\bibnamefont
  {Girard}}, \bibinfo {author} {\bibfnamefont {J.}~\bibnamefont {Lerm\'e}},
  \bibinfo {author} {\bibfnamefont {H.}~\bibnamefont {Gehan}}, \bibinfo
  {author} {\bibfnamefont {J.}~\bibnamefont {Margueritat}},\ and\ \bibinfo
  {author} {\bibfnamefont {A.}~\bibnamefont {Mermet}},\ }\bibfield  {title}
  {\bibinfo {title} {Mechanisms of resonant low frequency {R}aman scattering
  from metallic nanoparticle {L}amb modes},\ }\href
  {https://doi.org/10.1063/1.4983119} {\bibfield  {journal} {\bibinfo
  {journal} {J. Chem. Phys.}\ }\textbf {\bibinfo {volume} {146}},\ \bibinfo
  {pages} {194201} (\bibinfo {year} {2017})}\BibitemShut {NoStop}%
\bibitem [{\citenamefont {Girard}\ \emph {et~al.}(2019)\citenamefont {Girard},
  \citenamefont {Lerm{\'e}}, \citenamefont {Gehan}, \citenamefont {Mermet},
  \citenamefont {Bonnet}, \citenamefont {Cottancin}, \citenamefont {Crut},\
  and\ \citenamefont {Margueritat}}]{GirardJPCC19}%
  \BibitemOpen
  \bibfield  {author} {\bibinfo {author} {\bibfnamefont {A.}~\bibnamefont
  {Girard}}, \bibinfo {author} {\bibfnamefont {J.}~\bibnamefont {Lerm{\'e}}},
  \bibinfo {author} {\bibfnamefont {H.}~\bibnamefont {Gehan}}, \bibinfo
  {author} {\bibfnamefont {A.}~\bibnamefont {Mermet}}, \bibinfo {author}
  {\bibfnamefont {C.}~\bibnamefont {Bonnet}}, \bibinfo {author} {\bibfnamefont
  {E.}~\bibnamefont {Cottancin}}, \bibinfo {author} {\bibfnamefont
  {A.}~\bibnamefont {Crut}},\ and\ \bibinfo {author} {\bibfnamefont
  {J.}~\bibnamefont {Margueritat}},\ }\bibfield  {title} {\bibinfo {title}
  {Inelastic light scattering by multiple vibrational modes in individual gold
  nanodimers},\ }\href {https://doi.org/10.1021/acs.jpcc.9b03090} {\bibfield
  {journal} {\bibinfo  {journal} {J. Phys. Chem. C}\ }\textbf {\bibinfo
  {volume} {123}},\ \bibinfo {pages} {14834} (\bibinfo {year}
  {2019})}\BibitemShut {NoStop}%
\bibitem [{\citenamefont {Monta\~no Priede}\ \emph {et~al.}(2022)\citenamefont
  {Monta\~no Priede}, \citenamefont {Mlayah},\ and\ \citenamefont
  {Large}}]{MontaOPriedePRB22}%
  \BibitemOpen
  \bibfield  {author} {\bibinfo {author} {\bibfnamefont {J.~L.}\ \bibnamefont
  {Monta\~no Priede}}, \bibinfo {author} {\bibfnamefont {A.}~\bibnamefont
  {Mlayah}},\ and\ \bibinfo {author} {\bibfnamefont {N.}~\bibnamefont
  {Large}},\ }\bibfield  {title} {\bibinfo {title} {Raman energy density in the
  context of acoustoplasmonics},\ }\href
  {https://doi.org/10.1103/physrevb.106.165425} {\bibfield  {journal} {\bibinfo
   {journal} {Phys. Rev. B}\ }\textbf {\bibinfo {volume} {106}},\ \bibinfo
  {pages} {165425} (\bibinfo {year} {2022})}\BibitemShut {NoStop}%
\bibitem [{\citenamefont {Vasileiadis}\ \emph {et~al.}(2022)\citenamefont
  {Vasileiadis}, \citenamefont {Noual}, \citenamefont {Wang}, \citenamefont
  {Graczykowski}, \citenamefont {Djafari-Rouhani}, \citenamefont {Yang},\ and\
  \citenamefont {Fytas}}]{VasileiadisNL22}%
  \BibitemOpen
  \bibfield  {author} {\bibinfo {author} {\bibfnamefont {T.}~\bibnamefont
  {Vasileiadis}}, \bibinfo {author} {\bibfnamefont {A.}~\bibnamefont {Noual}},
  \bibinfo {author} {\bibfnamefont {Y.}~\bibnamefont {Wang}}, \bibinfo {author}
  {\bibfnamefont {B.}~\bibnamefont {Graczykowski}}, \bibinfo {author}
  {\bibfnamefont {B.}~\bibnamefont {Djafari-Rouhani}}, \bibinfo {author}
  {\bibfnamefont {S.}~\bibnamefont {Yang}},\ and\ \bibinfo {author}
  {\bibfnamefont {G.}~\bibnamefont {Fytas}},\ }\bibfield  {title} {\bibinfo
  {title} {Optomechanical hot-spots in metallic nanorod--polymer
  nanocomposites},\ }\href {https://doi.org/10.1021/acsnano.2c06673} {\bibfield
   {journal} {\bibinfo  {journal} {{ACS} Nano}\ }\textbf {\bibinfo {volume}
  {16}},\ \bibinfo {pages} {20419} (\bibinfo {year} {2022})}\BibitemShut
  {NoStop}%
\bibitem [{\citenamefont {Gelfand}\ and\ \citenamefont
  {Pelton}(2024)}]{GelfandJPCC24}%
  \BibitemOpen
  \bibfield  {author} {\bibinfo {author} {\bibfnamefont {R.}~\bibnamefont
  {Gelfand}}\ and\ \bibinfo {author} {\bibfnamefont {M.}~\bibnamefont
  {Pelton}},\ }\bibfield  {title} {\bibinfo {title} {Unified finite-element
  model for transient absorption and {R}aman scattering of vibrating noble
  metal nanoparticles},\ }\href {https://doi.org/10.1021/acs.jpcc.4c04071}
  {\bibfield  {journal} {\bibinfo  {journal} {J. Phys. Chem. C}\ }\textbf
  {\bibinfo {volume} {128}},\ \bibinfo {pages} {17526} (\bibinfo {year}
  {2024})}\BibitemShut {NoStop}%
\bibitem [{\citenamefont {Saison-Francioso}\ \emph {et~al.}(2020)\citenamefont
  {Saison-Francioso}, \citenamefont {L\'ev\^eque},\ and\ \citenamefont
  {Akjouj}}]{SaisonFranciosoJPCC20}%
  \BibitemOpen
  \bibfield  {author} {\bibinfo {author} {\bibfnamefont {O.}~\bibnamefont
  {Saison-Francioso}}, \bibinfo {author} {\bibfnamefont {G.}~\bibnamefont
  {L\'ev\^eque}},\ and\ \bibinfo {author} {\bibfnamefont {A.}~\bibnamefont
  {Akjouj}},\ }\bibfield  {title} {\bibinfo {title} {Numerical modeling of
  acousto--plasmonic coupling in metallic nanoparticles},\ }\href
  {https://doi.org/10.1021/acs.jpcc.0c00874} {\bibfield  {journal} {\bibinfo
  {journal} {J. Phys. Chem. C}\ }\textbf {\bibinfo {volume} {124}},\ \bibinfo
  {pages} {12120} (\bibinfo {year} {2020})}\BibitemShut {NoStop}%
\bibitem [{\citenamefont {Otomalo}\ \emph {et~al.}(2022)\citenamefont
  {Otomalo}, \citenamefont {Di~Mario}, \citenamefont {Hamon}, \citenamefont
  {Constantin}, \citenamefont {Toschi}, \citenamefont {Do}, \citenamefont
  {Juv\'e}, \citenamefont {Ruello}, \citenamefont {O'Keeffe}, \citenamefont
  {Catone}, \citenamefont {Paladini},\ and\ \citenamefont
  {Palpant}}]{OtomaloC22}%
  \BibitemOpen
  \bibfield  {author} {\bibinfo {author} {\bibfnamefont {T.~O.}\ \bibnamefont
  {Otomalo}}, \bibinfo {author} {\bibfnamefont {L.}~\bibnamefont {Di~Mario}},
  \bibinfo {author} {\bibfnamefont {C.}~\bibnamefont {Hamon}}, \bibinfo
  {author} {\bibfnamefont {D.}~\bibnamefont {Constantin}}, \bibinfo {author}
  {\bibfnamefont {F.}~\bibnamefont {Toschi}}, \bibinfo {author} {\bibfnamefont
  {K.-V.}\ \bibnamefont {Do}}, \bibinfo {author} {\bibfnamefont
  {V.}~\bibnamefont {Juv\'e}}, \bibinfo {author} {\bibfnamefont
  {P.}~\bibnamefont {Ruello}}, \bibinfo {author} {\bibfnamefont
  {P.}~\bibnamefont {O'Keeffe}}, \bibinfo {author} {\bibfnamefont
  {D.}~\bibnamefont {Catone}}, \bibinfo {author} {\bibfnamefont
  {A.}~\bibnamefont {Paladini}},\ and\ \bibinfo {author} {\bibfnamefont
  {B.}~\bibnamefont {Palpant}},\ }\bibfield  {title} {\bibinfo {title}
  {Acoustic vibration modes of gold--silver core--shell nanoparticles},\ }\href
  {https://doi.org/10.3390/chemosensors10050193} {\bibfield  {journal}
  {\bibinfo  {journal} {Chemosensors}\ }\textbf {\bibinfo {volume} {10}},\
  \bibinfo {pages} {193} (\bibinfo {year} {2022})}\BibitemShut {NoStop}%
\bibitem [{\citenamefont {Bragas}\ \emph {et~al.}(2023)\citenamefont {Bragas},
  \citenamefont {Maier}, \citenamefont {Boggiano}, \citenamefont {Grinblat},
  \citenamefont {Bert\'e}, \citenamefont {Menezes},\ and\ \citenamefont
  {Cort\'es}}]{BragasJOSAB23}%
  \BibitemOpen
  \bibfield  {author} {\bibinfo {author} {\bibfnamefont {A.~V.}\ \bibnamefont
  {Bragas}}, \bibinfo {author} {\bibfnamefont {S.~A.}\ \bibnamefont {Maier}},
  \bibinfo {author} {\bibfnamefont {H.~D.}\ \bibnamefont {Boggiano}}, \bibinfo
  {author} {\bibfnamefont {G.}~\bibnamefont {Grinblat}}, \bibinfo {author}
  {\bibfnamefont {R.}~\bibnamefont {Bert\'e}}, \bibinfo {author} {\bibfnamefont
  {L.~d.~S.}\ \bibnamefont {Menezes}},\ and\ \bibinfo {author} {\bibfnamefont
  {E.}~\bibnamefont {Cort\'es}},\ }\bibfield  {title} {\bibinfo {title}
  {Nanomechanics with plasmonic nanoantennas: ultrafast and local exchange
  between electromagnetic and mechanical energy},\ }\href
  {https://doi.org/10.1364/josab.482384} {\bibfield  {journal} {\bibinfo
  {journal} {J. Opt. Soc. Am. B}\ }\textbf {\bibinfo {volume} {40}},\ \bibinfo
  {pages} {1196} (\bibinfo {year} {2023})}\BibitemShut {NoStop}%
\bibitem [{\citenamefont {Timm}\ \emph {et~al.}(2022)\citenamefont {Timm},
  \citenamefont {Saviot}, \citenamefont {Crut}, \citenamefont {Blanchard},
  \citenamefont {Roiban}, \citenamefont {Masenelli-Varlot}, \citenamefont
  {Joly-Pottuz},\ and\ \citenamefont {Margueritat}}]{TimmJPCC22}%
  \BibitemOpen
  \bibfield  {author} {\bibinfo {author} {\bibfnamefont {M.~M.}\ \bibnamefont
  {Timm}}, \bibinfo {author} {\bibfnamefont {L.}~\bibnamefont {Saviot}},
  \bibinfo {author} {\bibfnamefont {A.}~\bibnamefont {Crut}}, \bibinfo {author}
  {\bibfnamefont {N.}~\bibnamefont {Blanchard}}, \bibinfo {author}
  {\bibfnamefont {L.}~\bibnamefont {Roiban}}, \bibinfo {author} {\bibfnamefont
  {K.}~\bibnamefont {Masenelli-Varlot}}, \bibinfo {author} {\bibfnamefont
  {L.}~\bibnamefont {Joly-Pottuz}},\ and\ \bibinfo {author} {\bibfnamefont
  {J.}~\bibnamefont {Margueritat}},\ }\bibfield  {title} {\bibinfo {title}
  {Study of single gold nanocrystals by inelastic light scattering
  spectroscopy},\ }\href {https://doi.org/10.1021/acs.jpcc.2c00077} {\bibfield
  {journal} {\bibinfo  {journal} {J. Phys. Chem. C}\ }\textbf {\bibinfo
  {volume} {126}},\ \bibinfo {pages} {3606} (\bibinfo {year}
  {2022})}\BibitemShut {NoStop}%
\bibitem [{\citenamefont {Vernier}\ \emph {et~al.}(2023)\citenamefont
  {Vernier}, \citenamefont {Saviot}, \citenamefont {Fan}, \citenamefont
  {Courty},\ and\ \citenamefont {Portal\`es}}]{VernierAN23}%
  \BibitemOpen
  \bibfield  {author} {\bibinfo {author} {\bibfnamefont {C.}~\bibnamefont
  {Vernier}}, \bibinfo {author} {\bibfnamefont {L.}~\bibnamefont {Saviot}},
  \bibinfo {author} {\bibfnamefont {Y.}~\bibnamefont {Fan}}, \bibinfo {author}
  {\bibfnamefont {A.}~\bibnamefont {Courty}},\ and\ \bibinfo {author}
  {\bibfnamefont {H.}~\bibnamefont {Portal\`es}},\ }\bibfield  {title}
  {\bibinfo {title} {Sensitivity of localized surface plasmon resonance and
  acoustic vibrations to edge rounding in silver nanocubes},\ }\href
  {https://doi.org/10.1021/acsnano.3c06990} {\bibfield  {journal} {\bibinfo
  {journal} {{ACS} Nano}\ }\textbf {\bibinfo {volume} {17}},\ \bibinfo {pages}
  {20462} (\bibinfo {year} {2023})}\BibitemShut {NoStop}%
\bibitem [{\citenamefont {Saviot}(2021)}]{SaviotN21}%
  \BibitemOpen
  \bibfield  {author} {\bibinfo {author} {\bibfnamefont {L.}~\bibnamefont
  {Saviot}},\ }\bibfield  {title} {\bibinfo {title} {Free vibrations of
  anisotropic nano-objects with rounded or sharp corners},\ }\href
  {https://doi.org/10.3390/nano11071838} {\bibfield  {journal} {\bibinfo
  {journal} {Nanomaterials}\ }\textbf {\bibinfo {volume} {11}},\ \bibinfo
  {pages} {1838} (\bibinfo {year} {2021})}\BibitemShut {NoStop}%
\bibitem [{\citenamefont {Portal\`es}\ \emph {et~al.}(2008)\citenamefont
  {Portal\`es}, \citenamefont {Goubet}, \citenamefont {Saviot}, \citenamefont
  {Adichtchev}, \citenamefont {Murray}, \citenamefont {Mermet}, \citenamefont
  {Duval},\ and\ \citenamefont {Pileni}}]{PortalesPNAS08}%
  \BibitemOpen
  \bibfield  {author} {\bibinfo {author} {\bibfnamefont {H.}~\bibnamefont
  {Portal\`es}}, \bibinfo {author} {\bibfnamefont {N.}~\bibnamefont {Goubet}},
  \bibinfo {author} {\bibfnamefont {L.}~\bibnamefont {Saviot}}, \bibinfo
  {author} {\bibfnamefont {S.}~\bibnamefont {Adichtchev}}, \bibinfo {author}
  {\bibfnamefont {D.~B.}\ \bibnamefont {Murray}}, \bibinfo {author}
  {\bibfnamefont {A.}~\bibnamefont {Mermet}}, \bibinfo {author} {\bibfnamefont
  {E.}~\bibnamefont {Duval}},\ and\ \bibinfo {author} {\bibfnamefont {M.-P.}\
  \bibnamefont {Pileni}},\ }\bibfield  {title} {\bibinfo {title} {Probing
  atomic ordering and multiple twinning in metal nanocrystals through their
  vibrations},\ }\href {https://doi.org/10.1073/pnas.0803748105} {\bibfield
  {journal} {\bibinfo  {journal} {Proc. Natl. Acad. Sci. U.S.A}\ }\textbf
  {\bibinfo {volume} {105}},\ \bibinfo {pages} {14784} (\bibinfo {year}
  {2008})}\BibitemShut {NoStop}%
\bibitem [{\citenamefont {Visscher}\ \emph {et~al.}(1991)\citenamefont
  {Visscher}, \citenamefont {Migliori}, \citenamefont {Bell},\ and\
  \citenamefont {Reinert}}]{Visscher1991}%
  \BibitemOpen
  \bibfield  {author} {\bibinfo {author} {\bibfnamefont {W.~M.}\ \bibnamefont
  {Visscher}}, \bibinfo {author} {\bibfnamefont {A.}~\bibnamefont {Migliori}},
  \bibinfo {author} {\bibfnamefont {T.~M.}\ \bibnamefont {Bell}},\ and\
  \bibinfo {author} {\bibfnamefont {R.~A.}\ \bibnamefont {Reinert}},\
  }\bibfield  {title} {\bibinfo {title} {On the normal modes of free vibration
  of inhomogeneous and anisotropic elastic objects},\ }\href
  {https://doi.org/10.1121/1.401643} {\bibfield  {journal} {\bibinfo  {journal}
  {J. Acoust. Soc. Am.}\ }\textbf {\bibinfo {volume} {90}},\ \bibinfo {pages}
  {2154} (\bibinfo {year} {1991})}\BibitemShut {NoStop}%
\bibitem [{\citenamefont {Mochizuki}(1987)}]{Mochizuki1987}%
  \BibitemOpen
  \bibfield  {author} {\bibinfo {author} {\bibfnamefont {E.}~\bibnamefont
  {Mochizuki}},\ }\bibfield  {title} {\bibinfo {title} {Application of group
  theory to free oscillations of an anisotropic rectangular parallelepiped.},\
  }\href {https://doi.org/10.4294/jpe1952.35.159} {\bibfield  {journal}
  {\bibinfo  {journal} {J. Phys. Earth}\ }\textbf {\bibinfo {volume} {35}},\
  \bibinfo {pages} {159} (\bibinfo {year} {1987})}\BibitemShut {NoStop}%
\bibitem [{\citenamefont {Pennec}\ \emph {et~al.}(2014)\citenamefont {Pennec},
  \citenamefont {Laude}, \citenamefont {Papanikolaou}, \citenamefont
  {Djafari-Rouhani}, \citenamefont {Oudich}, \citenamefont {El~Jallal},
  \citenamefont {Beugnot}, \citenamefont {Escalante},\ and\ \citenamefont
  {Mart{\'\i}nez}}]{PennecN14}%
  \BibitemOpen
  \bibfield  {author} {\bibinfo {author} {\bibfnamefont {Y.}~\bibnamefont
  {Pennec}}, \bibinfo {author} {\bibfnamefont {V.}~\bibnamefont {Laude}},
  \bibinfo {author} {\bibfnamefont {N.}~\bibnamefont {Papanikolaou}}, \bibinfo
  {author} {\bibfnamefont {B.}~\bibnamefont {Djafari-Rouhani}}, \bibinfo
  {author} {\bibfnamefont {M.}~\bibnamefont {Oudich}}, \bibinfo {author}
  {\bibfnamefont {S.}~\bibnamefont {El~Jallal}}, \bibinfo {author}
  {\bibfnamefont {J.~C.}\ \bibnamefont {Beugnot}}, \bibinfo {author}
  {\bibfnamefont {J.~M.}\ \bibnamefont {Escalante}},\ and\ \bibinfo {author}
  {\bibfnamefont {A.}~\bibnamefont {Mart{\'\i}nez}},\ }\bibfield  {title}
  {\bibinfo {title} {Modeling light-sound interaction in nanoscale cavities and
  waveguides},\ }\href {https://doi.org/10.1515/nanoph-2014-0004} {\bibfield
  {journal} {\bibinfo  {journal} {Nanophotonics}\ }\textbf {\bibinfo {volume}
  {3}},\ \bibinfo {pages} {413} (\bibinfo {year} {2014})}\BibitemShut {NoStop}%
\bibitem [{\citenamefont {Hecht}(2012)}]{freefem}%
  \BibitemOpen
  \bibfield  {author} {\bibinfo {author} {\bibfnamefont {F.}~\bibnamefont
  {Hecht}},\ }\bibfield  {title} {\bibinfo {title} {New development in
  {F}ree{F}em++},\ }\href {https://doi.org/10.1515/jnum-2012-0013} {\bibfield
  {journal} {\bibinfo  {journal} {J. Numer. Math.}\ }\textbf {\bibinfo {volume}
  {20}},\ \bibinfo {pages} {251} (\bibinfo {year} {2012})}\BibitemShut
  {NoStop}%
\bibitem [{\citenamefont {Murray}\ and\ \citenamefont
  {Saviot}(2004)}]{MurrayPRB04}%
  \BibitemOpen
  \bibfield  {author} {\bibinfo {author} {\bibfnamefont {D.~B.}\ \bibnamefont
  {Murray}}\ and\ \bibinfo {author} {\bibfnamefont {L.}~\bibnamefont
  {Saviot}},\ }\bibfield  {title} {\bibinfo {title} {Phonons in an
  inhomogeneous continuum: Vibrations of an embedded nanoparticle},\ }\href
  {https://doi.org/10.1103/PhysRevB.69.094305} {\bibfield  {journal} {\bibinfo
  {journal} {Phys. Rev. B}\ }\textbf {\bibinfo {volume} {69}},\ \bibinfo
  {pages} {094305} (\bibinfo {year} {2004})}\BibitemShut {NoStop}%
\bibitem [{\citenamefont {Laude}\ and\ \citenamefont
  {Wang}(2023)}]{LaudePRB2023}%
  \BibitemOpen
  \bibfield  {author} {\bibinfo {author} {\bibfnamefont {V.}~\bibnamefont
  {Laude}}\ and\ \bibinfo {author} {\bibfnamefont {Y.-F.}\ \bibnamefont
  {Wang}},\ }\bibfield  {title} {\bibinfo {title} {Quasinormal mode
  representation of radiating resonators in open phononic systems},\ }\href
  {https://doi.org/10.1103/PhysRevB.107.144301} {\bibfield  {journal} {\bibinfo
   {journal} {Phys. Rev. B}\ }\textbf {\bibinfo {volume} {107}},\ \bibinfo
  {pages} {144301} (\bibinfo {year} {2023})}\BibitemShut {NoStop}%
\bibitem [{\citenamefont {Geuzaine}\ and\ \citenamefont
  {Remacle}(2009)}]{gmsh}%
  \BibitemOpen
  \bibfield  {author} {\bibinfo {author} {\bibfnamefont {C.}~\bibnamefont
  {Geuzaine}}\ and\ \bibinfo {author} {\bibfnamefont {J.}~\bibnamefont
  {Remacle}},\ }\bibfield  {title} {\bibinfo {title} {Gmsh: A 3‐{D} finite
  element mesh generator with built‐in pre‐ and post‐processing
  facilities},\ }\href {https://doi.org/10.1002/nme.2579} {\bibfield  {journal}
  {\bibinfo  {journal} {Int. J. Num. Met. Eng.}\ }\textbf {\bibinfo {volume}
  {79}},\ \bibinfo {pages} {1309} (\bibinfo {year} {2009})}\BibitemShut
  {NoStop}%
\bibitem [{\citenamefont {Johnson}\ and\ \citenamefont
  {Christy}(1972)}]{JohnsonPRB72}%
  \BibitemOpen
  \bibfield  {author} {\bibinfo {author} {\bibfnamefont {P.~B.}\ \bibnamefont
  {Johnson}}\ and\ \bibinfo {author} {\bibfnamefont {R.~W.}\ \bibnamefont
  {Christy}},\ }\bibfield  {title} {\bibinfo {title} {Optical constants of the
  noble metals},\ }\href {https://doi.org/10.1103/physrevb.6.4370} {\bibfield
  {journal} {\bibinfo  {journal} {Phys. Rev. B}\ }\textbf {\bibinfo {volume}
  {6}},\ \bibinfo {pages} {4370} (\bibinfo {year} {1972})}\BibitemShut
  {NoStop}%
\bibitem [{\citenamefont {Jackson}(1999)}]{Jackson}%
  \BibitemOpen
  \bibfield  {author} {\bibinfo {author} {\bibfnamefont {J.~D.}\ \bibnamefont
  {Jackson}},\ }\href@noop {} {\emph {\bibinfo {title} {Classical
  electrodynamics}}},\ \bibinfo {edition} {3rd}\ ed.\ (\bibinfo  {publisher}
  {Wiley},\ \bibinfo {address} {New York, {NY}},\ \bibinfo {year} {1999})\
  \bibinfo {note} {see page 157}\BibitemShut {NoStop}%
\bibitem [{\citenamefont {Saviot}\ and\ \citenamefont
  {Murray}(2009)}]{SaviotPRB09}%
  \BibitemOpen
  \bibfield  {author} {\bibinfo {author} {\bibfnamefont {L.}~\bibnamefont
  {Saviot}}\ and\ \bibinfo {author} {\bibfnamefont {D.~B.}\ \bibnamefont
  {Murray}},\ }\bibfield  {title} {\bibinfo {title} {Acoustic vibrations of
  anisotropic nanoparticles},\ }\href
  {https://doi.org/10.1103/PhysRevB.79.214101} {\bibfield  {journal} {\bibinfo
  {journal} {Phys. Rev. B}\ }\textbf {\bibinfo {volume} {79}},\ \bibinfo
  {pages} {214101} (\bibinfo {year} {2009})}\BibitemShut {NoStop}%
\bibitem [{\citenamefont {Lamb}(1881)}]{LambPLMS1881}%
  \BibitemOpen
  \bibfield  {author} {\bibinfo {author} {\bibfnamefont {H.}~\bibnamefont
  {Lamb}},\ }\bibfield  {title} {\bibinfo {title} {On the vibrations of an
  elastic sphere},\ }\href {https://doi.org/10.1112/plms/s1-13.1.189}
  {\bibfield  {journal} {\bibinfo  {journal} {Proc. London Math. Soc.}\
  }\textbf {\bibinfo {volume} {s1-13}},\ \bibinfo {pages} {189} (\bibinfo
  {year} {1881})}\BibitemShut {NoStop}%
\bibitem [{\citenamefont {Kuok}\ \emph {et~al.}(2003)\citenamefont {Kuok},
  \citenamefont {Lim}, \citenamefont {Ng}, \citenamefont {Liu},\ and\
  \citenamefont {Wang}}]{KuokPRL03}%
  \BibitemOpen
  \bibfield  {author} {\bibinfo {author} {\bibfnamefont {M.~H.}\ \bibnamefont
  {Kuok}}, \bibinfo {author} {\bibfnamefont {H.~S.}\ \bibnamefont {Lim}},
  \bibinfo {author} {\bibfnamefont {S.~C.}\ \bibnamefont {Ng}}, \bibinfo
  {author} {\bibfnamefont {N.~N.}\ \bibnamefont {Liu}},\ and\ \bibinfo {author}
  {\bibfnamefont {Z.~K.}\ \bibnamefont {Wang}},\ }\bibfield  {title} {\bibinfo
  {title} {Brillouin study of the quantization of acoustic modes in
  nanospheres},\ }\href {https://doi.org/10.1103/physrevlett.90.255502}
  {\bibfield  {journal} {\bibinfo  {journal} {Phys. Rev. Lett.}\ }\textbf
  {\bibinfo {volume} {90}},\ \bibinfo {pages} {255502} (\bibinfo {year}
  {2003})}\BibitemShut {NoStop}%
\bibitem [{\citenamefont {Barron}\ \emph {et~al.}(2007)\citenamefont {Barron},
  \citenamefont {Zhu}, \citenamefont {Hecht}, \citenamefont {Tranter},\ and\
  \citenamefont {Isaacs}}]{BarronJMS07}%
  \BibitemOpen
  \bibfield  {author} {\bibinfo {author} {\bibfnamefont {L.~D.}\ \bibnamefont
  {Barron}}, \bibinfo {author} {\bibfnamefont {F.}~\bibnamefont {Zhu}},
  \bibinfo {author} {\bibfnamefont {L.}~\bibnamefont {Hecht}}, \bibinfo
  {author} {\bibfnamefont {G.~E.}\ \bibnamefont {Tranter}},\ and\ \bibinfo
  {author} {\bibfnamefont {N.~W.}\ \bibnamefont {Isaacs}},\ }\bibfield  {title}
  {\bibinfo {title} {Raman optical activity: An incisive probe of molecular
  chirality and biomolecular structure},\ }\href
  {https://doi.org/10.1016/j.molstruc.2006.10.033} {\bibfield  {journal}
  {\bibinfo  {journal} {J. Mol. Struct.}\ }\textbf {\bibinfo {volume}
  {834--836}},\ \bibinfo {pages} {7} (\bibinfo {year} {2007})}\BibitemShut
  {NoStop}%
\bibitem [{\citenamefont {Cho}\ \emph {et~al.}(2020)\citenamefont {Cho},
  \citenamefont {Byun}, \citenamefont {Lim}, \citenamefont {Im}, \citenamefont
  {Kim}, \citenamefont {Lee}, \citenamefont {Ahn},\ and\ \citenamefont
  {Nam}}]{ChoAN20}%
  \BibitemOpen
  \bibfield  {author} {\bibinfo {author} {\bibfnamefont {N.~H.}\ \bibnamefont
  {Cho}}, \bibinfo {author} {\bibfnamefont {G.~H.}\ \bibnamefont {Byun}},
  \bibinfo {author} {\bibfnamefont {Y.-C.}\ \bibnamefont {Lim}}, \bibinfo
  {author} {\bibfnamefont {S.~W.}\ \bibnamefont {Im}}, \bibinfo {author}
  {\bibfnamefont {H.}~\bibnamefont {Kim}}, \bibinfo {author} {\bibfnamefont
  {H.-E.}\ \bibnamefont {Lee}}, \bibinfo {author} {\bibfnamefont {H.-Y.}\
  \bibnamefont {Ahn}},\ and\ \bibinfo {author} {\bibfnamefont {K.~T.}\
  \bibnamefont {Nam}},\ }\bibfield  {title} {\bibinfo {title} {Uniform chiral
  gap synthesis for high dissymmetry factor in single plasmonic gold
  nanoparticle},\ }\href {https://doi.org/10.1021/acsnano.9b10094} {\bibfield
  {journal} {\bibinfo  {journal} {{ACS} Nano}\ }\textbf {\bibinfo {volume}
  {14}},\ \bibinfo {pages} {3595} (\bibinfo {year} {2020})}\BibitemShut
  {NoStop}%
\end{thebibliography}%
\end{document}